# Responsible AI Adoption in the Public Sector: A Data-Centric Taxonomy of AI Adoption Challenges

Anastasija Nikiforova
University of Tartu, Estonia
anastasija.nikiforova@ut.ee

Martin Lnenicka
University of Hradec Kralove, Czech Republic
martin.lnenicka@uhk.cz

Ulf Melin
Linkoping University, Sweden
ulf.melin@liu.se

David Valle-Cruz
UAEM, Mexico
davacr@uaemex.mx

Asif Gill
University of Technology Sydney, Australia
asif.gill@uts.edu.au

Cesar Casiano Flores
University of Twente, the Netherlands
c.a.casianoflores@utwente.nl

Emyana Sirait
TU Delft, The Netherlands
e.r.e.sirait@tudelft.nl

Mariusz Luterek
University of Warsaw, Poland
m.luterek@uw.edu.pl

Richard Michael Dreyling
Fortinet, USA
rmdreyling@gmail.com

Barbora Tesarova
University of Hradec Kralove, Czech Republic
barbora.tesarova@uhk.cz

**Abstract:** *Despite Artificial Intelligence (AI) transformative potential for public sector services, decision-making, and administrative efficiency, adoption remains uneven due to complex technical, organizational, and institutional challenges. Responsible AI frameworks emphasize fairness, accountability, and transparency, aligning with principles of trustworthy AI and fair AI, yet remain largely aspirational, overlooking technical and institutional realities, especially foundational data and governance. This study addresses this gap by developing a taxonomy of data-related challenges to responsible AI adoption in government. Based on a systematic review of 43 studies and 21 expert evaluations, the taxonomy identifies 13 key challenges across technological, organizational, and environmental dimensions, including poor data quality, limited AI-ready infrastructure, weak governance, misalignment in human-AI decision-making, economic and environmental sustainability concerns. Annotated with institutional pressures, the taxonomy serves as a diagnostic tool to surface "symptoms" of high-risk AI deployment and guides policymakers in building the institutional and data governance conditions necessary for responsible AI adoption.*

**Keywords:** Artificial Intelligence, Data governance, Public Sector, Responsible AI, Technology Adoption.

## 1. Introduction

Artificial Intelligence (AI) is increasingly viewed as a cornerstone of digital transformation in the public sector, with the potential to innovate service delivery, automate workflows, and enhance decision-making (Farrell et al., 2023; Hernandez Quiros et al., 2022). Yet adoption remains uneven and fraught with risks, hindered by a complex web of challenges. A critical, often underexamined barrier lies in the interdependence between AI and public sector data. As the maxim "*garbage in, garbage out*" suggests, AI performance depends on high-quality, ethically sourced, and interoperable data. Yet many governments operate with siloed systems, inconsistent standards, and weak data governance, undermining privacy, fairness, explainability, and accountability (Balahur et al., 2022; Giest & Klievink, 2024; Kokina et al., 2025). In this sense, *data is the lifeblood of AI*, and without addressing its limitations, responsible adoption remains unattainable. While AI can sometimes help alleviate data challenges, it also introduces new risks, including algorithmic opacity, ethical uncertainty, and institutional strain (Díaz-Rodríguez et al., 2023; De Bruijn et al., 2022). This interdependence gives rise to what we term *data-AI* challenges - barriers that emerge at the intersection of AI deployment and Public Data Ecosystems (PDEs), often multidimensional, institutionally embedded, and difficult to resolve in isolation. Despite their centrality, these challenges are inconsistently defined and under-theorized, limiting both scholarly synthesis and practical guidance, especially for public administrators navigating complex implementation contexts (Farrell et al., 2023; Misuraca et al., 2020).

While interest in AI risks and *responsible AI* has grown substantially (Alshahrani et al., 2022; Madan & Ashok, 2023; Papagiannidis et al., 2023; Valle-Cruz et al., 2024), often framed under notions of *trustworthy AI* and *fair AI*, existing research tends to focus narrowly on technical, ethical, or organizational aspects. With the emerging focus on sustainability, environmental considerations associated with *sustainable AI* and *green AI* also start gaining attention. These debates are important, yet they consider the institutional and data governance preconditions necessary for translating such principles into practice. These studies tend to examine AI governance, transparency, or fairness in fragmented, domain-specific ways, typically without a unifying conceptual framework. This fragmentation limits theoretical synthesis and cross-case learning across the information systems and public administration research (Kuziemski & Misuraca, 2020; Zuiderwijk et al., 2021).

A further tension exists between the normative ambitions of responsible AI frameworks and the technical realities of contemporary AI systems. Established principles, such as explainability, fairness, or human control, are often difficult or impossible to fully realize due to the complexity, opacity, and scale of modern AI models (Mikalef et al., 2025). Consequently, many frameworks that are aspirational rather than operational. Compounding this issue, the responsible AI discourse frequently underplays the foundational role of data, even though robust data ecosystems are a prerequisite for operationalizing the high-level principles outlined in the EU High-Level Expert Group's *Ethics Guidelines for Trustworthy AI* (2019), the OECD *AI Principles* (2019), and UNESCO's *Recommendation on the Ethics of AI* (2021).

This study takes a different stance - rather than asking *whether AI systems themselves are responsible*, we focus on the manner of adoption, i.e., *whether public administrations are positioned to adopt AI responsibly,* i.e., in ways that uphold legality, accountability, fairness, sustainability, and inclusion. Responsible adoption is thus framed as a governance process rather than a system property. Our contribution is therefore not an additional ethical framework, but a taxonomy of *data–AI challenges* that systematizes the barriers shaping adoption. By identifying the institutional and data preconditions essential for responsible use, the taxonomy highlights the conditions whose absence can render AI adoption unlawful, opaque, unfair, unsustainable, or exclusionary.

The taxonomy is grounded in a *Systematic Literature Review* (SLR) and refined through expert validation with 21 experts. It is structured using the *Technology–Organization–Environment* (TOE) framework and enriched by *Institutional Theory* (DiMaggio & Powell, 1983; Currie, 2009), enabling a refined understanding of both structural and institutional dynamics shaping responsible AI adoption. Rather than offering a prescriptive model, the taxonomy serves as a flexible diagnostic tool for scholars and practitioners to analyze, compare, and prioritize challenges across public contexts. It provides a foundation for future empirical research, policy design, strategic planning.

This study makes three contributions: (1) addresses the underexamined role of data in responsible AI adoption by developing a taxonomy; (2) integrates TOE and Institutional Theory to uncover not only structural categories but also the institutional forces sustaining these challenges; (3) provides a practical yet flexible diagnostic tool, surfacing the data–institutional preconditions whose absence can result in unlawful, opaque, unfair, unsustainable, or exclusionary AI use.

The remainder of the paper is structured as follows: Section 2 provides the background and theoretical framing. Section 3 outlines the research methodology. Section 4 presents the taxonomy, and Section 5 discusses implications, limitations, and directions for future research. Conclusions are drawn in Section 6.

## 2. Background

### 2.1. The nature of Data-AI challenges

The adoption of AI in government is inseparable from the condition of PDEs. We define *data–AI challenges* as systemic frictions that arise at the intersection of AI systems and public data infrastructures. They are not only technical issues but multidimensional, socio-technical, and institutionally embedded barriers that directly shape whether AI can be adopted responsibly (Janssen et al., 2012; Veale & Brass, 2019; Tangi et al., 2025). These challenges manifest along several interrelated dimensions:

- data availability and quality – data are often siloed, incomplete, or inconsistent, limiting training, accuracy, and fairness (Janssen et al., 2012; Balahur et al., 2022; Farrell et al., 2023);
- governance and sharing – decentralized ownership, weak cross-agency coordination, legal ambiguity restrict lawful and effective data reuse (Dawes et al., 2016; Kuziemski & Misuraca, 2020; Lis & Otto, 2021);
- infrastructure and interoperability – outdated IT systems, poor metadata, and low interoperability hinder data integration and scalability (Desouza et al., 2020; Neumann et al., 2024; Gill & Nikiforova, 2025);
- human capacity and institutional culture – skills gaps, risk-averse organizational logics, and fragmented procurement constrain responsible adoption (Alshahrani et al., 2022; Campion et al., 2022; Misra et al., 2023);
- fairness, accountability, and ethics – biased or incomplete datasets risk discrimination in sensitive domains, compounded by institutional demands for explainability, transparency, and ethical safeguards (Wirtz et al., 2019; Mehrabi et al., 2021; De Bruijn et al., 2022; Kokina et al., 2025; Díaz-Rodríguez et al., 2023).

Together, they reflect long-standing data governance shortcomings amplified by AI's demands for explainability, integration, and oversight. They underscore that AI adoption challenges are structural and institutional, not just technical, requiring systematic conceptualization.

## 2.2. Theoretical framing

The *TOE* framework (Tornatzky & Fleischer, 1990) provides a foundation for analyzing technological innovation adoption across three dimensions: technology, organization, and environment. The *technological* dimension includes the internal and external infrastructure, tools, data quality, and capabilities that can enable or restrict the use of new technologies (e.g., AI). The *organizational* dimension covers characteristics of the entity, including its size, structure, leadership, culture, human capital, and resource availability, which shape the capacity for change. These dimensions determine an organization's readiness to take risks, invest in innovation, and manage technological change. The *environmental* dimension encompasses external pressures such as regulatory frameworks, competition, professional norms, inter-organizational dynamics, and public policy influences, which influence organizations' technological decisions (Oliveira & Martins, 2011). While TOE is relevant in the public sector, where AI adoption depends on internal capacity and alignment with external mandates and societal expectations (endogenous and exogenous drivers of change) (Oliveira & Martins, 2011), it is limited in explaining why certain challenges endure despite available technical solutions or institutional awareness.

Institutions are composed of regulative, normative, and cultural-cognitive elements that establish what is considered legitimate or rational behavior within an organizational field. These structures, though stable, can vary over time through processes of socialization, (re)production, and norm sedimentation (Björck, 2004). In this regard, to enhance the explanatory depth of our framework, we draw on *Institutional Theory* (DiMaggio & Powell, 1983), which allows us to analyze how organizational behavior is shaped not only by efficiency but also by legitimacy within an institutional field. DiMaggio and Powell (1983) identify three types of institutional isomorphism that help explain the persistence and spread of practices across organizations: (1) *coercive pressures*, which stem from formal legal, regulatory, or financial mandates (e.g., GDPR, procurement rules), (2) *normative pressures*, which emerge from shared professional ethics and norms (e.g., fairness, transparency, public accountability, auditability), (3) *mimetic pressures*, which refers to imitation of perceived best practices in other agencies or jurisdictions, often under conditions of uncertainty. These pressures have the potential to shape organizational decisions regarding AI adoption. Institutional Theory thus complements the TOE by helping to explain why certain barriers to AI adoption persist— not only due to structural constraints but also because of institutional dynamics such as compliance, imitation, or alignment with shared values. Together, these frameworks offer an integrated lens to identify both structural and institutional roots of data-related challenges in public sector AI adoption, informing the taxonomy, and supporting a more holistic understanding of what shapes responsible, sustainable deployment. This combined framework also guide the classification and interpretation of challenges identified through our literature review and expert evaluation (Section 3). Consistent with our abductive analytical strategy (Alvesson et al., 2022), TOE and Institutional Theory were not applied as predefined coding frames but used as interpretive lenses to structure the taxonomy after inductive challenge identification.

## 3. Methodology

To develop a taxonomy, we followed a structured methodology combining a SLR and expert validation. This process was guided by the taxonomy development method proposed by Nickerson et al. (2013), which offers a systematic approach for identifying and organizing dimensions (characteristics) and their respective values (types or categories) within a set of conceptually related phenomena. Rather than prescribing a fixed framework, the taxonomy is designed as a flexible analytical structure to support systemic thinking about the presence, absence, and interaction of data-related challenges across public sector contexts.

### 3.1. Systematic Literature Review

To ground the taxonomy in the existing body of knowledge, we conducted a SLR specifically focused on challenges at the intersection of AI and public sector data ecosystems, following the guidelines of Kitchenham et al. (2009). To this end, we queried digital libraries covered by Scopus and Clarivate Web of Science, complemented by Google Scholar, thereby ensuring a broad and peer-reviewed collection of academic literature. The search query was defined as a combination of terms "*public sector*", "*data*" and "*information*", *AI*, and "*challenge*", where the latter has been provided with synonyms. As such, the defined search string was: *TITLE-ABS-KEY(("public sector" OR "public administration") AND ("data" OR "information") AND ("AI" OR "Artificial Intelligence")) AND ALL(challenge OR barrier OR resistance factor OR problem OR obstacle).* The former part of the search string was applied to Title, Abstract and Keywords, whereas the latter, concerning challenges, was applied to all text. This strategy was used to ensure that our corpus consists of articles, where AI and public sector are the primary research objects mentioned in the title, abstract and keywords. In contrast, challenges surrounding this topic may be mentioned in the text, not mandatory being the central object of research, but an implication or observation made during research. Only English-language journal articles, conference papers, book chapters, and short surveys and letters published between 2010 and 2024 were considered.

This resulted in 272 papers in Scopus and 46 in Web of Science (as of September 2024). After deduplication, two reviewers screened the titles and abstracts of all 213 papers against pre-defined inclusion and exclusion criteria, focusing on relevance to AI in the public sector, data-related challenges. Discrepancies were resolved through discussion or consultation with a third reviewer, leading to the elimination of 164 papers outside the study's scope. Subsequently, full texts were retrieved and subjected to a more detailed assessment of their contributions to understanding data challenges in AI-driven ecosystems. Again, disagreements were resolved through consensus or third-party arbitration, resulting in a final sample of 43 papers for further in-depth analysis (Fig. 1) (the list of selected papers is available on Zenodo - https://doi.org/10.5281/zenodo.15613840).

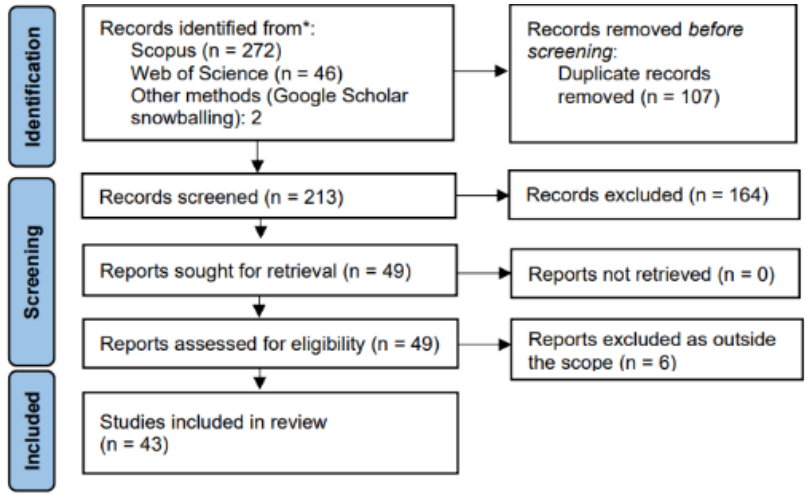

**Figure 1. PRISMA of the conducted SLR.**

The analysis of these 43 papers followed the *taxonomy development method* proposed by Nickerson et al. (2013), employing an inductive -empirical-to-conceptual- approach. This involved iterative cycles of data extraction, concept identification, and taxonomy refinement. Specifically, for each paper, relevant passages discussing data-related challenges were extracted and coded. These codes were grouped into higher-level concepts, which were further refined and organized into a hierarchical taxonomy. This resulted in a structured taxonomy capturing the most prominent, recurring, and theoretically significant categories of data-AI challenges reported in the literature.

To ensure completeness, consistency, and validity, the taxonomy development followed a structured, iterative process. This included (1) defining the *meta-characteristic* - challenges to AI adoption in the public sector, (2) identifying the intended *end users* - policymakers, public sector organizations, researchers, (3) selecting the approach to guide classification - *empirical-to-conceptual*. *Dimensions*, i.e., broad categories of challenges and their respective *characteristics* (i.e., specific sub-types) were defined iteratively. *Stopping conditions* were based on (a) mutual exclusivity, i.e., each challenge clearly belongs to one category, (b) collective exhaustiveness, i.e., no major challenges remain unclassified, and (c) expert validation, i.e., no new dimensions emerged from literature or expert feedback. The resulting taxonomy was then *evaluated* by experts to assess for duplication, ambiguity, or overly broad categories, ensuring each entry was well-defined and practically relevant.

### 3.2. Expert validation and refinement

To validate and refine the preliminary list, we performed *a multi-expert evaluation*. Expert selection employed a purposive sampling strategy, designed to identify individuals possessing expertise directly relevant to the study's objectives. We targeted individuals with demonstrable proficiency in data management or AI within the context of the public sector, coupled with substantial knowledge in the second domain. Experts were required to hold a minimum of a master's degree in a discipline, such as Information Systems, Public Administration, Business and Management, or Social Sciences, and demonstrate a minimum of five years of professional experience in a related field. To mitigate potential local biases and ensure the generalizability of findings across digital governance contexts, the sample encompassed experts from a range of countries, including those exhibiting varying levels of digital government maturity.

Considering the exploratory nature of this task with the need to capture diverse insights and ensure the robustness of the initial taxonomy, we targeted 15 to 25 participants. This range aligns with recommendations in the literature, which suggest 10–20 participants for such studies (Hsu & Sandford, 2007). After approaching over 30 identified candidates, 21 agreed to participate, forming our expert panel. The expert panel represented a diverse range of expertise with backgrounds in Computer Science (Information Systems, Database Systems, Software Engineering), Public Administration (Governance of Urban Technologies and Sustainability, Public Policy), and interdisciplinary fields combining these disciplines, based in Europe, e.g., Austria, Czech Republic, Estonia, Germany, Netherlands, Poland, Spain, Sweden but also Americas, e.g., Brazil, Mexico or United States. The panel included academics holding positions from Junior Research Fellow to Full Professor, holding concurrent or prior professional roles in the public sectors, including data analysts, IT consultants, system analysts/designers, and managers of data centers and ICT programs within government organizations. Their expertise spanned various aspects of the data roles and data lifecycle. Several experts had direct experience with AI adoption in public services, data initiatives, and the governance of PDEs, ensuring a multifaceted perspective on the data challenges associated with AI in the public sector. We asked experts to assess the list qualitatively, providing comments: (1) to improve or reformulate specific challenges; (2) to merge overlapping items; (3) for overlooked challenges; (4) on challenges considered less relevant or unsuitable. This led to the initial version of the taxonomy.

### 3.3. Taxonomy structuring

Our approach follows a hybrid abductive strategy, beginning inductively with the systematic review and subsequently using TOE and Institutional Theory as sensitizing lenses to iteratively structure and interpret the taxonomy. Since we seek a comprehensive list of challenges associated with AI adoption, it is expected to include multidimensional, complex challenges with socio-political, technical, and institutional nuances. Thus, we structure the refined version of the taxonomy using the *TOE* (Tornatzky & Fleischer, 1990) framework as the main theoretical lens to organize challenges across systemic domains, thereby facilitating more meaningful analysis of categories within it. This is informed by the conceptual discussion in Section 2.2.

To complement the structural classification, we apply Institutional Theory (DiMaggio & Powell, 1983) as an analytical lens, identifying whether each challenge was primarily shaped by coercive, normative, or mimetic pressures. This interpretive step is carried out collaboratively among the authors to ensure consistency and theoretical alignment. The goal of this dual-theoretical integration is not only to classify the challenges structurally using TOE but also to reveal how institutional environments shape the conditions under which AI is adopted in the public sector. The final taxonomy thus presents both structural and institutional layers of analysis, offering a more comprehensive understanding of the challenges emerging in AI implementation. This approach aligns with prior information systems and public administration research that emphasizes the influence of institutional forces in technology adoption (e.g., Weerakkody et al., 2009).

## 4. Taxonomy of Data-AI challenges to responsible AI adoption

This section presents the taxonomy of data-related challenges to responsible AI adoption in the public sector. The taxonomy is informed by a SLR, refined through expert validation, and structured using the TOE and Institutional Theory lenses (Section 2 and 3).

The initial taxonomy is derived from a systematic review of 43 academic studies. 18 challenges were extracted, frequently cited in AI and public sector literature, covering a range of interrelated issues (Table 1, with full references and descriptions in Zenodo).

Key challenges include availability and integration of data sources (Campion et al., 2022; Wirtz et al., 2019), poor data quality and metadata standards (Misra et al., 2023; Tan, 2023), algorithm opacity (Sun & Medaglia, 2019), infrastructure constraints such as computing power and storage (Ojo et al., 2019; Alshahrani et al., 2022), weak governance and unclear accountability in data ecosystems (Pautz, 2023), human-AI misalignment and overload (Madan & Ashok, 2023), and limited data and AI literacy among staff and citizens (Neumann et al., 2024; Zuiderwijk et al., 2021). Ethical and institutional issues such as privacy risks, fairness concerns, and compliance with AI regulations feature strongly as well (Campion et al., 2022; Van Noordt et al., 2025). Several studies also highlight budgetary constraints (Sun & Medaglia, 2019). Collectively, these challenges reflect the multifaceted barriers to responsible AI adoption.

Table 1. Challenges identified in the literature.

| Challenges | Short description |
| --- | --- |
| Availability of data sources | Difficulties in identifying and accessing suitable and sufficient data for AI |
| Data integration | Fragmented, heterogeneous, and inaccessible data across systems |
| AI-ready data | Poor data quality, lack of metadata, bias, and standardization issues |
| Data algorithms | Algorithm opacity, lack of interpretability, and standard frameworks |
| Data timeliness | Outdated or infrequently updated datasets |
| Computing resources | Limited processing power and outdated infrastructure |
| Data storage resources | Insufficient and outdated storage infrastructure |
| Governance and management | Lack of coordination, strategy, and governance for data and AI |
| External data sharing | Resistance to inter-agency collaboration and legal/policy barriers |
| Data interpretation | Risks of bias, discrimination, and misinterpretation of AI-generated outputs |
| Data security and privacy | Cybersecurity gaps and insufficient privacy protection frameworks |
| Ethics and trust | Concerns around fairness, legitimacy, and social acceptance of AI |
| Accountability and transparency | Ambiguity over responsibility, explainability, and rules for AI use |
| Legal and administrative barriers | Regulatory uncertainty and risk-averse cultures in the public sector |
| Costs and financial constraints | High AI costs, unclear return on investment, and limited budgets |
| Human-AI decision-making | Misalignment between AI logic and human judgment, including information overload |

To refine the initial list, 21 domain experts reviewed these challenges. Based on expert input, several refinements were made: (1) items were merged (e.g., *Computing resources*, *Data storage*, and *Data integration* were unified under *Data infrastructure*); (2) themes were reorganized with challenges related to *ethics, explainability*, and *accountability* combined under *Impacts on decision-making and use*; (3) new categories - *Data and AI systems, platforms, tools, and services*, and dimensions of *economic* and *environmental sustainability* were added; (4) descriptions were expanded, with categories such as *AI-ready data* now explicitly including aspects of *provenance*, while *Stakeholders, data and AI literacy* emphasizes *capabilities*, *inclusion*, and *continuous upskilling*. Rather than removing any challenges, concerns around redundancy or generality were addressed through rewording and thematic integration. This iterative process resulted in a more coherent taxonomy of 13 challenges (Table 2, column 2 and 3).

To structure our taxonomy and enhance analytical depth, we conducted a theoretical mapping process using TOE lenses to categorize the refined list of challenges. *Technological* challenges include limited access to suitable data sources, outdated computing infrastructure, and rigid, non-interoperable AI systems. Poor data quality and data AI-readiness, incl. data incompleteness, bias, lack of metadata, along with the absence of algorithmic transparency or standards, further constrain responsible AI deployment in the complex public sector settings. *Organizational* challenges involve weak governance structures, unclear responsibilities, and limited strategic planning for AI. Misalignment between human-AI decision-making, accountability gaps, and low data-AI literacy among stakeholders further hinder responsible adoption in public institutions. *Environmental* challenges reflect external pressures that shape AI adoption in the public sector. These include resistance to inter-agency data sharing, legal and policy constraints, weak coordination mechanisms, cybersecurity vulnerabilities, and underdeveloped ethical governance. Financial sustainability is also a barrier, with high implementation costs, limited returns, and constrained public budgets, compounded by often-overlooked environmental impacts such as AI's energy consumption and carbon footprint. TOE allows for overlapping categorization, hence we introduce *cross-cutting challenges* category, however, keep the dominant logic for simplification with challenges placed under their primary or most directly related domain. Cross-cutting challenges bridge organizational and environmental domains. They include systemic concerns around data and AI security, privacy, and ethics, such as bias, transparency, and public trust, that require both internal safeguards and compliance with external legal, cultural, and societal norms. We find these issues foundational to responsible AI that cannot be addressed in isolation from broader institutional and governance dynamics.

Table 2. Taxonomy of Data-AI challenges to responsible AI adoption

| TOE dimension | Challenge | Description | Institutional Pressure |
|---|---|---|---|
| Technology | Availability of data sources | Difficulty in identifying, accessing, and evaluating suitable and sufficient data sources for AI use incl. fragmented ownership, missing datasets, and unknown availability | Coercive (regulatory data mandates) Normative (data quality standards) |
| | Computing and data storage resources | Limited computing power and storage capacities, outdated infrastructure, insufficient scalability to support AI deployment, and frequent upgrade requirements | Coercive (funding/regulation on IT standards) Normative (industry best practices) |
| | Data and AI systems, platforms, tools, and services | Challenges in acquiring, developing, and maintaining modern, interoperable systems and services tailored for AI use, incl. system rigidity, vendor lock-in, and lack of modular tools | Mimetic (following successful vendors/peers) Normative (technical standards) |
| | AI-ready data | Poor data quality (e.g., inconsistency, incompleteness), lack of metadata and provenance, absence of standard formats, bias in data, and challenges with entity reconciliation | Normative (professional data ethics) Coercive (data compliance laws) |
| | Algorithm transparency and standardization | Opacity of algorithms, lack of standard design principles or benchmarking frameworks, inability to process unstructured data, and concerns around misuse and fairness | Coercive (regulations on transparency), Normative (ethical algorithm design), Mimetic (peer benchmarking) |
| Organization | Governance and management of data, AI systems, and PDEs | Weak internal coordination, unclear process responsibilities, and lack of sustainable strategies for long-term AI governance | Coercive (internal compliance) Normative (organizational culture) Mimetic (benchmarking governance models) |
| | Actions and interactions between humans, data, and AI | Misalignment between human judgment and machine logic, challenges in human-in-the-loop systems, and information overload | Normative (professional roles/values) Mimetic (adopting best practices) |
| | Impacts on decision-making and use | Uncertainty over the fairness, accuracy, or reliability of AI-generated decisions, accountability gaps in decision support | Coercive (accountability laws) Normative (organizational ethics) Mimetic (following peer policies) |
| | Stakeholders, data, and AI literacy | Lack of awareness, knowledge, and technical competence among public officials and other stakeholders involved in AI use | Normative (professional education standards) Coercive (training mandates) |
| Environment | Data collaboration and exchange | Resistance to inter-organizational data sharing, poor coordination between agencies, misalignment of goals or expectations, and legal or policy barriers | Coercive (legal data sharing mandates) Normative (sector norms) Mimetic (peer collaboration trends) |
| | Data and AI security, privacy, and ethics | Inadequate cybersecurity measures, risks to citizen privacy, and a lack of frameworks to ensure AI safety and ethical use | Coercive (privacy laws) Normative (ethical expectations) Mimetic (industry norms) |
| | Economic sustainability | High implementation and maintenance costs, uncertain financial returns and long-term benefits, and budgetary constraints for AI adoption | Coercive (budget policies) Normative (financial best practices) |
| | Environmental sustainability | Energy consumption and carbon footprint related to training and operating AI models, often overlooked in public contexts | Normative (environmental responsibility) Coercive (regulatory requirements) |
| Cross-cutting *(O+E)* | Data and AI security, privacy, and ethics | Challenges related to ensuring cybersecurity and protecting sensitive data, alongside ethical considerations such as fairness, bias mitigation, transparency, and public trust, spanning internal organizational safeguards compliance with external legal, social, and cultural norms | Coercive (laws/regulations) Normative (ethics codes) Mimetic (peer standards) |
| | Impacts on decision-making and use | The influence of AI technologies on organizational workflows and human decision-making processes, incl. challenges of trust, acceptance, and the transformation of roles | Normative (social acceptance) Mimetic (organizational imitation) |
| | Stakeholders, data and AI literacy | Gaps in knowledge, skills, and competencies regarding AI and data among public officials and citizens, incl. challenges in continuous training, addressing digital divides, promoting inclusion, and fostering understanding of AI's benefits and risks at organizational and societal levels. | Normative (education norms) Coercive (training mandates) |

In addition to TOE classification, each challenge was annotated with the dominant institutional pressures influencing its emergence or persistence: *coercive pressures* are seen to be especially influential in shaping data availability, infrastructure, accountability, and compliance with transparency or privacy laws; *normative pressures* are prominent in challenges related to AI readiness, decision-making, and stakeholder literacy; *mimetic pressures* are particularly visible in the adoption of tools, governance models, and human-AI collaboration strategies. Most challenges are influenced by multiple overlapping institutional forces, emphasizing that (public sector) AI adoption is not just a technical or managerial process but an institutionally embedded transformation shaped by legitimacy concerns, external expectations, and established norms. This mapping provides clarification on why certain challenges persist despite available solutions and points to deeper reforms needed across structural and normative dimensions. The mapping was validated during expert review sessions and is presented alongside the TOE classification in the final taxonomy in Table 2.

## 5. Discussion

This study presents a theory-informed taxonomy that systematically classifies the data-related challenges of AI adoption in the public sector. Grounded in the TOE and Institutional Theory, the taxonomy reveals how technical, organizational, and environmental factors interact with institutional pressures to shape implementation barriers. It moves beyond technical limitations to expose the socio-political, normative, and capacity-related dynamics at play in public AI deployments. The taxonomy is agnostic to adoption intent and applies to AI adoption in general, with its contribution to *responsible* adoption lying in making visible the preconditions and trade-offs that determine if adoption will be lawful, accountable, fair, sustainable, and inclusive. In this sense, the taxonomy complements ongoing debates on *trustworthy AI, fair AI, and sustainable* or *green AI*. While these discourses articulate high-level principles, our contribution is to diagnose the institutional and data governance conditions that must be in place for such principles to be actionable in the public sector. By surfacing where these preconditions are absent, the taxonomy helps policymakers and practitioners understand why ambitions of trustworthiness, fairness, or sustainability often remain aspirational in practice.

### 5.1. Implications for research and practice

This study holds *implications for research* by advancing digital government, public administration, and IS scholarship on responsible AI through a flexible, theory-informed lens for analyzing data-related challenges across contexts. Its dual grounding in TOE and Institutional Theory supports comparative work on how technical and institutional barriers manifest and persist. Rather than a fixed roadmap, the taxonomy serves as a scaffold for systemic and context-sensitive analysis, particularly in areas of data governance.

It also lays a foundation for empirical studies, such as case research or surveys that can assess the relative importance of these challenges across public data roles (e.g., users, producers, regulators) and settings. Furthermore, its structure supports longitudinal studies on how barriers evolve as AI maturity and institutional readiness develop, contributing to a more nuanced understanding of AI governance in the public sector. At the same time, the taxonomy also serves for comparative analysis across jurisdictions or sectors, enabling the identification of consistent patterns and critical gaps in both institutional readiness and data infrastructure concerning AI adoption.

Beyond its role in academic classification, the framework offers a *practical* diagnostic instrument for public administrators, helping them pinpoint systemic obstacles, prioritize interventions, and align AI initiatives with existing organizational capabilities and diverse stakeholder needs. To support this, government enterprise architecture frameworks can be leveraged to responsibly address and integrate the data challenges inherent in AI adoption within public sector processes (Kamalabai et al., 2024).

In addition, the taxonomy encourages practitioners to move beyond purely technical solutions toward holistic strategies that embed governance, collaboration, and ethical data stewardship in the design and implementation of AI initiatives (Kuziemski & Misuraca, 2020; De Bruijn et al., 2022).

While grounded in the public sector, many of the identified challenges (data quality, infrastructure readiness, algorithmic transparency, and sustainability) are domain-agnostic, making the taxonomy a useful reference point for organizations more broadly.

### 5.2. Limitations

This study has several limitations. The SLR was limited by the quality and scope of academic literature, potentially overrepresenting conceptual contributions while underrepresenting practical implementations or unintended consequences of AI use and missing relevant grey literature and non-English studies.

Although expert validation strengthened the findings, the taxonomy may not encompass all contextual nuances across public sector domains and national contexts. Its scope is limited to what could be captured through the literature review and expert validation, and further empirical testing is needed to generalize its applicability.

Driven by the idea of responsible AI adoption as a governance approach rather than attributes of the AI model, the taxonomy identifies broad challenges in public sector AI adoption, influenced by institutional norms and expectations, rather than the specific normative commitments typically associated with responsible AI. Consequently, it has limited connections to common responsibility commitments, which limits its immediate use as a diagnostic tool for public administrations. Future empirical testing and refinement are needed to address these gaps.

## 6. Conclusion

This study developed a taxonomy of data–AI challenges that shape governments' ability to adopt AI responsibly. By combining the TOE with Institutional Theory, the taxonomy highlights that barriers are not simply technical but rooted in organizational capacity and institutional logics. This study makes three main contributions. Theoretically, it advances digital government and IS research by conceptualizing *data–AI challenges* as socio-technical and institutionally embedded barriers, grounded in the TOE and Institutional Theory frameworks. Methodologically, it develops a taxonomy through a SLR and expert validation, offering a scaffold for comparative and longitudinal studies. Practically, it provides a diagnostic tool for policymakers and administrators to identify and address preconditions for adopting AI responsibly.

In this study we reframe responsibility - rather than asking *whether AI systems are responsible by design*, we ask *whether public administrations are positioned to adopt AI responsibly*. This shift directs attention to data readiness, governance arrangements, and institutional accountability as preconditions for responsible adoption. More broadly, the study underscores a tension in current debates, i.e., while responsible AI frameworks articulate high-level values, they risk remaining aspirational unless connected to the realities of data ecosystems and public institutions. Our taxonomy offers one step toward bridging this gap by systematizing the practical barriers that determine whether such values can be realized in government contexts.

Future research will build on this foundation by prioritizing challenges across data lifecycle stages and enhancing the taxonomy with a "*responsibility overlay*," mapping each challenge to responsibility commitments (e.g., privacy, accountability, fairness, sustainability, inclusion). This will provide more fine-grained guidance for policymakers and practitioners and strengthen the taxonomy's value as a practical diagnostic tool, supporting a pragmatic approach to responsible AI adoption in the public sector.

## 7. Acknowledgements

This research was (co-)funded by the Excellence project, the Faculty of Informatics and Management, University of Hradec Kralove, Czech Republic.